\documentstyle[twocolumn,aps,prb]{revtex}

\tighten

\begin{document}

\title{Linear response conductance and magneto-resistance of
ferromagnetic single-electron transistors}

\author{Arne Brataas} \address{Lyman Laboratory of Physics, Harvard
University, Cambridge, MA 02138}

\author{X. H. Wang} \address{Department of Physics, Lund University,
S{\"o}lvegatan 14A, S-223 62 Lund, Sweden}

\date{\today}
\maketitle

\begin{abstract}
The current through ferromagnetic single-electron transistors (SET's)
is considered. Using path integrals the linear response conductance is
formulated as a function of the tunnel conductance vs. quantum
conductance and the temperature vs. Coulomb charging energy. The
magneto-resistance of ferromagnet-normal metal-ferromagnet (F-N-F)
SET's is almost independent of the Coulomb charging energy and is only
reduced when the transport dwell time is longer than the spin-flip
relaxation time. In all-ferromagnetic (F-F-F) SET's with negligible
spin-flip relaxation time the magneto-resistance is calculated
analytically at high temperatures and numerically at low
temperatures. The F-F-F magneto-resistance is enhanced by higher order
tunneling processes at low temperatures in the 'off' state when the
induced charges vanishes. In contrast, in the 'on' state near
resonance the magneto-resistance ratio is a non-monotonic function of
the inverse temperature.
\end{abstract}

\pacs{PACS numbers:  75.70.Pa, 73.40.Gk, 73.23.Hk}

\section{Introduction}

The spin of the electron can provide new functionality in electronic
devices\cite{Prinz98:282} and spin-polarized transport is therefore an
active research field.  Spin-transport was pioneered by Tedrow and
Meservey\cite{Meservey94:173} and the discovery of the giant
magnetoresistance effect (GMR) in metallic multilayers has motivated a
substantial re-newed interest in this field. Recently spin-polarized
current injection in all-semiconductor devices\cite{Fiederling99:787}
and carbon nanotubes\cite{Tsukagoshi99:572} have been realized. The
fabrication of tunnel junctions made of two ferromagnetic leads
separated by an insulating layer has become well under
control.\cite{Moodera95:3273,Ono97:1261} These technological
advancements can make it possible to design new generation of
electronic devices such as magnetic RAM's, sensors and ultimately
perhaps quantum computers.

In this article, we will focus on an elementary structure, the
ferromagnetic single-electron transistor and discuss the role of the
ferromagnetic electrodes, spin-accumulation and Coulomb charging
effects on the transport properties.  A schematic picture of the
system under consideration is shown in Fig.\ \ref{f:set}. A normal
(ferromagnetic) metal island is coupled to two ferromagnetic
reservoirs by tunnel junctions. There is a capacative coupling $C$
between the metal island and the metal reservoirs. The island is so
small so that the charging energy $E_c=e^2/2C$ associated with the
addition of a single electron to the island can be larger than both
the temperature and the applied source-drain bias. The electrostatic
potential on the island can be controlled by a gate voltage $V_g$
capacitively coupled via a capacitance $C_g$ to the island. We
consider an island weakly coupled to the reservoirs by many channels
with transmission probabilities much less than unity, but the total
tunnel conductance $G$ can be smaller or larger than the quantum
conductance $G_K$. In the simplest case, the single-electron
transistor (SET) comprising of tunnel junctions with junction
resistances much larger than the quantum resistance $R_K=h/e^2$
exhibits Coulomb blockade of the single electron tunneling at low
temperatures when the gate voltage vanishes, and a conductance peak
when $C_g V_g=e/2$.  For spin-polarized charge transfer the different
transport properties of the spin-up and spin-down electrons have to be
taken into account. Therefore the behavior of the SET is determined by
the interplay between spin-dependent transport and Coulomb charging
effects.
\cite{Barnas98:1058,Takahashi98:1758,Barnas98:85,Brataas99:93,Korotkov99:89,Imamura99:6017,Wang99:5138}

All-ferromagnetic (F-F-F) SET's have recently been investigated
experimentally \cite{Ono97:1261} and
theoretically.\cite{Barnas98:1058,Takahashi98:1758,Wang99:5138} The
tunneling magneto-resistance (TMR) in F-F-F systems is defined as the
relative conductance difference by changing the magnetization of the
ferromagnetic island from being parallel to anti-parallel to the
magnetizations of the ferromagnetic reservoirs (which are kept
parallel). Ferromagnetic islands have a short spin-flip relaxation
time and several works have computed the transport properties of F-F-F
systems disregarding
spin-accumulation.\cite{Barnas98:1058,Takahashi98:1758,Wang99:5138} An
enhanced TMR in the Coulomb blockade regime was found
experimentally,\cite{Ono97:1261} and it was sub-sequentially shown
that the TMR could be doubled due to co-tunneling
\cite{Takahashi98:1758} and that higher order tunneling processes
could give even larger enhancements of the TMR.\cite{Wang99:5138} In
what follows, we will use a non-perturbative approach to calculate the
tunneling magneto-resistance (TMR) of the F-F-F SET showing that even
larger enhancements of the TMR than what has been reported in Ref.\
\onlinecite{Wang99:5138} is possible. We will also discuss the TMR in
the 'on'-state when the Coulomb blockade effect is lifted. It will be
demonstrated that the TMR in this regime is a non-monotonic function
of the temperature.

The physics of the TMR in F-N-F systems is completely
different from the mechanism of the TMR in F-F-F 
systems. F-F-F systems exhibit a finite TMR even in the
absence of any spin-accumulation on the ferromagnetic island. The
transport properties can be modeled by an all-normal metal device with
magnetization configuration dependent tunnel conductances.  In
contrast, the TMR of F-N-F systems is uniquely related
to the spin-accumulation on the island. Spin-accumulation governs the
transport properties and leads to a measurable TMR when
the spin-flip relaxation time $\tau_{\text{sf}}$ is longer than the
dwell time so that \cite{Brataas99:93}
\begin{equation}
\tau_{\text{sf}}/(\chi h)>G_K/G \, ,
\label{spincondition}
\end{equation}
where $\chi$ is the spin-susceptibility, $G$ is the tunnel conductance
and $G_K=e^2/h$ is the quantum conductance.  In order to realize
spin-accumulation a small spin-susceptibility is required
(\ref{spincondition}) which implies a small normal metal island with a
large Coulomb charging energy $E_c$ and consequently Coulomb blockade
effects should be taken into account.\cite{Brataas99:93} When the
tunnel conductance $G$ is much smaller than the quantum conductance
$G_K$ and the temperature is not too low, transport can be described
by sequential tunneling processes and this was performed in Refs.\
\onlinecite{Barnas98:85,Brataas99:93,Korotkov99:89}. However, Eq.\
(\ref{spincondition}) shows that spin-accumulation is easier to
realize when the tunnel conductance is large, $G > G_K$. In this
strong tunneling limit the quantum fluctuations of the charge on the
island as well as the spin-acccumulation should be taken into account
with a theoretical description that goes beyond the low order
perturbative treatment presented in Refs.\
\onlinecite{Barnas98:85,Brataas99:93,Korotkov99:89,Imamura99:6017}. The
tunneling magneto-resistance (TMR) in F-N-F systems is defined in a
different way than for (F-F-F) systems: The TMR is the relative
conductance difference on going from a configuration where the
magnetizations in the ferromagnetic reservoirs are parallel to a
configuration where the magnetizations are anti-parallel
$(g_{\text{P}}-g_{\text{AP}})/g_{\text{P}}$. Changing the
magnetization configuration in F-N-F systems only changes the
spin-dependence of the tunnel conductances but conserves the total
tunnel conductances.  Hence, a much weaker coupling to the charge
degrees of freedom of the island and a reduced influence of Coulomb
charging should be expected. In the Coulomb blockade regime at low
temperatures the total conductance through the system is
suppressed. Consequently, also the transport of spins into the island
decreases and at sufficiently low temperatures the dwell time is
longer than the spin-flip relaxation time so that
(\ref{spincondition}) is no longer satisfied. The spin-accumulation
and thus the TMR must therefore be reduced in the Coulomb blockade
regime which is contrary to the behavior of the TMR in F-F-F
systems. We will show that the linear response TMR in F-N-F systems in
the Coulomb blockade regime equals or is smaller than the classical
high-temperature TMR, which is in contrast to the results for F-F-F
systems.

We present in this paper a unified analytical expression for the
linear response conductance of F-F-F and F-N-F SET's with dirty
metallic islands. Our formula is valid to all orders of the ratio
between the tunnel conductances and the quantum conductance, for
arbitrary spin-flip relaxation times and takes into account general
band-structures for the ferromagnetic reservoirs and the
normal/ferromagnetic metal island. Our results are in this sense
general. The analytical expression for the linear response conductance
is used to discuss the magneto-resistance of F-F-F and F-N-F SET's.

The paper is organized in the following way.  The system is described
and the model Hamiltonian introduced in Section \ref{s:Model}.  In
Section \ref{s:lincond} the linear response conductance is calculated,
including the multi-band effect of the electrodes and the
spin-accumulation on the central island.  The TMR is evaluated in
Section \ref{f:FNF} in the case of ferromagnet-normal
metal-ferromagnet SET's and in Section \ref{s:FFF} in the case of
all-ferromagnet SET's. Section \ref{s:conclusion} concludes the paper.

\section{Model}
\label{s:Model}

We consider a ferromagnetic single-electron transistor comprising of a
normal/ferromagnetic metal island connected to ferromagnetic leads by
tunnel junctions.  The central island is capacitively coupled to a
gate voltage $V_g$ via a capacitance $C_g$. The Hamiltonian of such a
device is
\begin{equation}
\hat{H} = \left( \hat{H}_{l}+\hat{H}_{i}+\hat{H}_{r} \right) +
\hat{H}_c + \left( \hat{H}_{li}+\hat{H}_{ri}+\text{h.c.} \right) \, .
\end{equation}
The quasi-particles in the left reservoir are described by
\begin{equation}
\hat{H}_{l} =\sum_{{\bf l} \sigma nm}W_{{\bf l}\sigma }^{nm}
\hat{c}_{{\bf l}\sigma }^{n\dag } \hat{c}_{{\bf l}\sigma }^{m} \, , 
\end{equation}
where $n$ and $m$ are the band-indices, $\sigma$ is the electron spin,
${\bf l}$ is the momentum in the left reservoir, $W^{nm}_{{\bf l}
\sigma}$ determine the quasi-particle energy bands and the
hybridization between the bands for a given spin, and $\hat{c}_{{\bf
l}}^{n\dag}$ creates an electron with spin $\sigma$ and momentum ${\bf
l}$ in band $n$ in the left reservoir. The hat ($\hat{}$) denotes an
operator. There are similar Hamiltonians for the quasi-particles on
the island ($l \rightarrow i$) and the quasi-particles in the right
reservoir ($l \rightarrow r$).  The Coulomb charging effects are
included via
\begin{equation}
\hat{H}_{c} = \frac{e^{2}}{2C}\left( \sum_{\sigma} \hat{N}^i_{\sigma}
-n_{\text{ex}} \right) ^{2} \, ,
\end{equation}
where the induced electron number is $n_{\text{ex}}=C_g V_g/e$ and the
spin-dependent number of excess electrons on the island is
$\hat{N}^i_{\sigma}=\sum_{{\bf i}n} \hat{c}_{{\bf i}\sigma}^{n\dagger}
\hat{c}_{{\bf i}\sigma}^n$. We will below also make use of the
spin-dependent excess number of electrons in the left (right)
reservoir, $\hat{N}^l_{\sigma}=\sum_{{\bf l}n} \hat{c}_{{\bf
l}\sigma}^{n\dagger} \hat{c}_{{\bf l}\sigma}^n$
($\hat{N}^r_{\sigma}=\sum_{{\bf r}n} \hat{c}_{{\bf
r}\sigma}^{n\dagger} \hat{c}_{{\bf r}\sigma}^n$).  The tunneling
between the reservoirs and the island is taken into account via the
tunneling Hamiltonians $\hat{H}_{pi}=\sum_{\sigma} \hat{H}_{pi\sigma}$
($p=l$ denotes left and $p=r$ denotes right):
\begin{equation}
\hat{H}_{pi\sigma } =\sum_{{\bf pi}nm}t_{{\bf pi}\sigma }^{nm}
\hat{c}_{{\bf p}\sigma }^{n\dag } \hat{c}_{{\bf i}\sigma }^{m} \, . \\
\end{equation}	
The tunneling matrix elements $t_{{\bf pi}\sigma}^{nm}$ allow
tunneling between different bands with the same spin.  Spin-flip
scatterings during the tunneling processes are disregarded, but can be
included in the formalism leading to a renormalization of the
spin-asymmetries of the junction resistances.

We consider the situation when the temperature is much larger than the
level spacing and the level spacing is much smaller than the Coulomb
charging energy. The level spacing is consequently taken to be
continous. Co-tunneling and higher order tunneling processes have both
elastic and inelastic contributions,\cite{Averin90:2446} e.g. in the
case of co-tunneling the elastic contribution means that the same
electron tunnels through both of the two junctions whereas the
inelastic contribution corresponds to two different electrons that
tunnel in the two junctions: One tunnels into the island above its
Fermi level, and another jumps out of the island from below the Fermi
level. The elastic contribution is sensitive to the phase of the
electron on the island and consequently becomes vanishingly small for
dirty systems when the characteristic time of tunneling through the
macroscopic barrier caused by the Coulomb energy $\hbar/E_c$
($E_c=e^2/(2C)$) is larger than the classical time $L^2/D$ for
diffusion through the island ($L$ is the size of the island and $D$ is
the diffusion coefficient).\cite{Averin90:2446} We consider the
inelastic transport regime exclusively and disregard the elastic
contributions to the tunneling processes, i.e. for an estimate of a
diffusion constant on the island of $D=10cm^2/s$ and a charging energy
of $E_c=1 K$ we assume that the size of the island $L < 0.1 \mu m$ (in
reality there is also scattering at the boundary of the island that
further relax this assumption). The occupation of the energy levels on
the island can then be described by an energy $\epsilon$ and spin
$\sigma$ dependent non-equilibrium distribution
$f_{\sigma}^{i}(\epsilon)$. The non-equilibrium distribution
$f_{\sigma}^i(\epsilon)$ is spin-dependent to allow a non-equilibrium
spin-accumulation and only equals the equilibrium Fermi-Dirac
distribution $f(\epsilon-\mu)$ when the bias voltage vanishes ($\mu$
is the equlibrium chemical potential). Furthermore, the leads and the
island are metallic so that the tunnel conductances (see Eq.\
(\ref{gdd_multi}) below) are energy independent on the scale of the
Coulomb charging energy which is much smaller than the Fermi
energy. We consider the linear response regime where the bias voltage
is the smallest energy scale in the system. Under these assumptions,
only the total occupation of spin-up and spin-down electrons on the
island, $\int d\epsilon [ f_{\sigma}^{i}-f(\epsilon-\mu)]$, determine
the macroscopic Coulomb barrier and the total effective tunneling
rates. Consequently to all orders in perturbation theory of the
tunneling Hamiltonian, we can describe the electron occupation on the
island by a local chemical potential $\mu^{i}_{\sigma} \equiv \mu +
\int d\epsilon [f_{\sigma}^i(\epsilon)-f(\epsilon-\mu)]$ since the
energy-dependence of the distribution $f_{\sigma}^i(\epsilon)$ does
not influence the transport properties. It is thus e.g. irrelevant if
there is substantial energy-relaxation on the island so that the
non-equilibrium distribution $f_{\sigma}^i(\epsilon)$ approaches the
Fermi-Dirac distribution $f(\epsilon-\mu_{\sigma}^i)$ or if there is
no energy-relaxation so that the non-equilibrium distribution
$f_{\sigma}^i(\epsilon)$ is a linear combination of the Fermi-Dirac
distributions in the left and right reservoirs with different local
chemical potentials.

The electrons in the right reservoir and in the left reservoir are in
local equilibrium with chemical potentials $\mu^{l}(t)$ and
$\mu^{r}(t)$, respectively and as seen above the electrons on the
island can be described as in local equilibrium with a spin-dependent
chemical potential $\mu^i_{\sigma}(t)$.  The temperature is the same in
all subsystems. The thermal average of the spin-dependent current from
the left ($p=l$) or right ($p=r$) reservoir to the island can then be
written as
\begin{equation}\label{gc}
I_{\sigma }^p(t) = ie \text{Tr}\left[ \left( \hat{H}_{pi\sigma
}(t)-\hat{H}_{pi\sigma }^{\dagger }(t)\right) \exp (-\beta
K(t))\right] \\
\end{equation}
with the non-equilibrium time-dependent grand canonical potential
\begin{equation}
\hat{K}(t)=\hat{H}-\sum_{\sigma }[\mu^{l}(t) \hat{N}^l_{\sigma
}+\mu^{r}(t) \hat{N}^r_{\sigma }+\mu^i_{\sigma }(t) \hat{N}^i_{\sigma
}] \,.
\label{grand_noneq}
\end{equation}
The spin-dependent chemical potentials on the island
$\mu^i_{\uparrow}$ and $\mu^i_{\downarrow}$ should be determined from
the flux of particles and spins into the island. The spin-flux into
the island is\cite{Brataas99:93}
\begin{equation}
(I_{\uparrow}^l+I_{\uparrow}^r)-(I_{\downarrow}^l+I_{\downarrow}^r) =
\frac{es}{\tau_{\text{sf}}} \, .
\label{spinbalance}
\end{equation}
Current conservation through the system requires
\begin{equation}
(I_{\uparrow}^l+I_{\downarrow}^l)+(I_{\uparrow}^r+I_{\downarrow}^r)=0
\, .
\label{curcons}
\end{equation}
The excess number of spins on the island $s$ is related to the
non-equilibrium difference in the chemical potentials by the
spin-susceptibility $\chi_s$ via $s=\chi_s
(\mu^i_{\uparrow}-\mu^i_{\downarrow})$.\cite{Brataas99:93,MacDonald9912391}
For non-interacting electrons $\chi_s=D$, where $D$ is the density of
states. The expression for the current (\ref{gc}) and the conservation
laws for spins (\ref{spinbalance}) and particles (\ref{curcons})
uniquely determine the spin-dependent local chemical potentials on the
island, $\mu_{i\uparrow}$ and $\mu_{i\downarrow}$, and consequently
the current through the device. We will focus on the linear response
regime in what follows.

\section{Linear response conductance}
\label{s:lincond}

In the linear response regime a perturbation expansion in terms of the
small differences in the chemical potentials on the left reservoir, on
the island and on the right reservoir can be performed: $\mu^{l}(t)
=\mu +\delta \mu^{l}(t)$, $\mu^{r}(t) =\mu +\delta \mu^{r}(t)$, and
$\mu^i_{\sigma}(t) =\mu +\delta \mu^i_{\sigma }(t)$.
The non-equilibrium grand canonical potential is
$\hat{K}(t)=\hat{K}_0+\delta \hat{K}(t)$, where the time-independent
unperturbed grand canonical potential is $\hat{K}_0=\hat{H}-\mu
\sum_{\sigma } ( \hat{N}^l_{\sigma }+\hat{N}^r_{\sigma
}+\hat{N}^i_{\sigma } ) $ and the time-dependent perturbation is
$\delta \hat{K}=-\sum_{\sigma } ( \delta \mu^{l} \hat{N}^l_{\sigma
}+\delta \mu^{r} \hat{N}^r_{\sigma }+\delta \mu^i_{\sigma }
\hat{N}^i_{\sigma } ) $. We perform a time-dependent unitary
transformation with the unitary matrix $\hat{U}(t) =\exp [i
\int_{-\infty }^{t}dt_{1}\delta \hat{K}(t_{1})]$. The transformed grand
canonical potential becomes $\hat{K}_t(t) = \hat{K}_0 + \delta
\hat{H}(t) $, where to the lowest order in the non-equilibrium
chemical potential differences
\begin{equation}
\delta \hat{H}(t)=\frac{1}{e} \sum_{qs} \hat{I}^q_{s}(t)
\int_{-\infty}^{t} dt_1 \left[ \delta \mu^q(t_1) - \delta
\mu^i_{s}(t_1) \right] 
\label{delH}
\end{equation}
and the sum over the parameter $q$ corresponds to the left reservoir
$q=l$ and the right reservoir $q=r$.  The spin-currents through the
left and the right barriers can be found from a standard linear
response calculation:
\begin{equation} \label{current}
I^p_{\sigma }(\omega ) = \frac{i}{e \omega } \sum_{qs}
\Pi^{pq}_{\sigma s}(\omega) \left[ \delta \mu^{q}(\omega) - \delta
\mu^i_s(\omega) \right] \, ,
\end{equation}
where the retarded current-current correlation function is
\begin{equation}
\Pi^{pq}_{\sigma s}(t) =-i \left\langle \left[ \hat{I}^p_{\sigma
}(t),\hat{I}^q_s(0)\right] \right\rangle _{\text{eq}} \, ,
\end{equation}
This correlation function will be evaluated using path integrals. The
computation is similar to the case of ultrasmall tunnel junctions or
metallic SET that has been studied
extensively.\cite{Schon90:198,Wang97:12868} The main steps in the
calculation of the correlation function is as
follows.\cite{Schon90:198,Wang97:12868} We use the equilibrium
finite-temperature Matsubara formalism to find the retarded
current-current correlation function.  The imaginary time generating
functional is defined as
\begin{equation}
Z[\left\{ \eta \right\} ]=\int D[\left\{ c^{\ast },c \right\} ]
D[\varphi ]e^{-S_g} \, ,
\end{equation}
with the generalized action 
\begin{eqnarray}
S_g &=& - \int_{0}^{\beta }d\tau
\text{Tr}_{{\bf p}n\sigma} c^{\ast }(\tau ) \partial_{\tau}
c(\tau ) \nonumber \\
&& +\int_0^{\beta} d\tau K_0(\tau ) - \int_0^{\beta} d\tau
\sum_{p\sigma} I^p_{\sigma }(\tau )\eta^p_{\sigma }(\tau ) \, .
\label{genaction}
\end{eqnarray}
A Hubbard-Stratonovich transformation is performed to reduce the
quadratic term in the charging energy to a linear coupling of the
particle number operator with an auxiliary field.\cite{Schon90:198}
After integrating out the electronic degrees of freedom and including
the principal contribution of the tunneling processes the generalized
effective action is derived. The current-current correlation function
can then be found by performing the functional derivative
$\Pi^{pq}_{\sigma s }(\tau, \tau')=Z^{-1} \partial^2 Z/(\partial
\eta^p_{\sigma }(\tau )\partial \eta^q _{s }(\tau' )) |_{\eta
\rightarrow 0} $.  The resulting Fourier transform of the
current-current correlation function is
\begin{equation} \label{ccc}
\frac{1}{\omega} \Pi^{pq}_{\sigma s}=
\frac{e^2}{h} \left[ \delta _{p\sigma ,qs} g^p_{\sigma }g_0 + 2
g^p_{\sigma } g^q_{s}g_1 \right] \, .
\end{equation}
The dimensionless tunnel conductance ($p=l,r$) is
\begin{equation} \label{gdd}
g^p_{ \sigma} = \sum_{jj^{\prime }}D_{p\sigma }^{j}D_{i\sigma
}^{j^{\prime }}\left| \tilde{t}_{p\sigma }^{jj^{\prime }}\right| ^{2}
\, ,
\label{gdd_multi}
\end{equation}
Here $D_{l\sigma}^{j}$, $D_{r\sigma }^{j}$ and $D_{i\sigma}^{j}$ are
the spin and band dependent densities of states and the renormalized
transmission coefficient at the Fermi energy is $ \tilde{t}_{p\sigma }
=\left(U_{p\sigma }\right) ^{\dagger }t_{p\sigma }U_{i\sigma }$, where
$U_{p\sigma}$ ($U_{i\sigma }$) is the material-specific matrix that
diagonalizes the band Hamiltonian in reservoir $p$ (island) for
electrons with spin $\sigma$.  In a single-band model the conductance
is simply proportional to the density of states in the reservoir and
the island, $g^p_{\sigma}=D_{p\sigma} D_{i\sigma} |t_{\sigma}|^2$.
The relation is in general more complicated as shown in
(\ref{gdd_multi}). The correlation function $g_0(i\omega _{n})$ is
\begin{equation}
g_0 =\frac{4\pi}{i\omega_n} \int d\tau e^{i\omega_n \tau } \chi (\tau
) \Gamma (\tau )
\label{g0}
\end{equation}
with the phase-phase correlation function 
\begin{eqnarray} \label{ppc}
\Gamma (\tau ) & = & \frac{1}{Z} \sum_{k=-\infty}^{\infty} e^{2\pi i k
n_{\text{ex}}} \times \nonumber \\
&& \int_{b_k} D[\varphi ] e^{-S[\varphi]} \cos \left[ \varphi (\tau
)-\varphi (0)\right] \, .
\label{phasephase}
\end{eqnarray}
The effective action is
\begin{eqnarray}
S[\varphi] & = & \int d\tau \frac{\left[ \dot{ \varphi}(\tau )\right]
^{2}}{4E_{c}} \nonumber \\
&& -g^{\text{cl}} \int \! \! d\tau \int \! \! d\tau' \chi \left( \tau
-\tau ^{\prime }\right) \cos \left[ \varphi (\tau) -\varphi (\tau
^{\prime })\right] \, .
\label{wirk}
\end{eqnarray}
with the boundary condition $b_{k}\rightarrow \varphi (\beta )=\varphi
(0)+2\pi k$.  Since the leads and island are (ferro-)metallic with
continuous spectra, the damping kernel $\chi (\tau )$ is an even
function of the imaginary time $\tau$ with a period $\beta $ and the
Matsubara-Fourier components $\chi (\omega _{l})=-|\omega_{l}|/4\pi $.
The coupling of the island to the reservoirs is determined by the
algebraic total dimensionless classical conductance
$g^{\text{cl}}=\sum_{p\sigma } g^p_{\sigma }$. The partition function
is $Z=\sum_{k} \int_{b_k} D[\varphi] \exp({-S[\varphi]})$.  Another
correlation function in (\ref{ccc}) is
\begin{eqnarray}
g_1 & = &\frac{4\pi}{i\omega_n} \int d\tau e^{i\omega_{n}\tau }
\frac{1}{Z} \times \nonumber \\
&& \sum_{k=-\infty}^{\infty} \! e^{2\pi i k n_{\text{ex}}} \int_{b_k} \!
D[\varphi ] e^{-S[\varphi]} I_{t}(\varphi ,\tau )I_{t}(\varphi ,0)
\label{g1}
\end{eqnarray}
where 
\begin{equation}
I_{t}(\varphi ,\sigma )=\int d\tau \chi \left( \sigma -\tau \right)
\sin \left[ \varphi (\sigma )-\varphi (\tau )\right] \,.
\end{equation}
Due to the current conservation (\ref{curcons}) the correlation
function $g_1$ (\ref{g1}) does not appear in the final form of the
stationary current.  

Knowing the current-current correlation function, we can find the
non-equilibrium chemical potential on the island by using the
conservation laws for spins (\ref{spinbalance}) and particles
(\ref{curcons}).  We find the spin-accumulation
\begin{equation}
\frac{\mu^i_{\uparrow}-\mu^i_{\downarrow}}{\mu^l-\mu^r}=
\frac{(g^l_{\uparrow}g^r_{\downarrow}-g^l_{\downarrow}
g^r_{\uparrow})/(g_{\uparrow}g_{\downarrow})}{1+g_{\text{sf}} \left[
(g_0 g_{\uparrow})^{-1}+(g_0 g_{\downarrow})^{-1} \right]}
\end{equation}
and the average chemical potential
\begin{equation}
\frac{\mu^i_{\uparrow}+\mu^i_{\downarrow}}{2}=
\frac{g^l \mu^l+g^r \mu^r}{g^l+g^r}
-\frac{g_{\uparrow}-g_{\downarrow}}{g_{\uparrow}+g_{\downarrow}}
 \frac{\mu^i_{\uparrow}-\mu^i_{\downarrow}}{2} \, ,
\end{equation}
where the spin-flip conductance $g_{\text{sf}}$ is related to the
spin-susceptibility $\chi_s$ and the spin-flip relaxation time
$\tau_{\text{sf}}$ via\cite{Brataas99:93} $g_{\text{sf}}=\pi \chi_s
/\tau_{\text{sf}}$. From the non-equilibrium chemical potentials on
the island we obtain the current through the systems by using the
relation for the current (\ref{current}) and finally we calculate the
conductance $g=eI/(\delta \mu^l - \delta \mu^r)$: The linear
conductance in units of the quantum conductance $e^2/h$ of the FSET is
\begin{equation}
g = \frac{g^r g^l}{g^r + g^l} g_0
\left[1-\frac{(g^l_{\downarrow} g^r_{\uparrow}-g^l_{\uparrow}
g^r_{\downarrow})^2/(g^rg^lg_{\uparrow}g_{\downarrow})}{1+g_{\text{sf}}\left[
(g_0g_{\uparrow})^{-1}+(g_0g_{\downarrow})^{-1} \right]} \right]
\label{cond} \, ,
\end{equation}
where $g^l_{\uparrow}$ and $g^l_{\downarrow}$ ($g^r_{\uparrow}$ and
$g^r_{\downarrow}$) are the spin-up and spin-down conductances of the
left (right) junction in units of the quantum conductance $e^2/h$, and
we have also introduced the combined conductances
$g^l=g^l_{\uparrow}+g^l_{\downarrow}$,
$g^r=g^r_{\uparrow}+g^r_{\downarrow}$,
$g_{\uparrow}=g^l_{\uparrow}+g^r_{\uparrow}$, and
$g_{\downarrow}=g^l_{\downarrow}+g^r_{\downarrow}$. The retarded
correlation function $g_0$ (\ref{g0}) is the same as for an all-normal
metal single-electron transistor. It renormalizes the conductance due
to the Coulomb charging effects and depends on the induced electron
number $C_g V_g/e$, the ratio of the charging energy to the
temperature $\beta E_c$, and the coupling between the island and the
reservoirs through the total conductance $g^r+g^l$.  Through Eq.\
(\ref{cond}) we have thus reduced the problem of finding the
conductance of ferromagnetic single-electron transistors to a
well-studied problem of the renormalized conductance in all-normal
metallic single-electron transistors. The second term in the bracket
of (\ref{cond}) is a signature of the spin-accumulation and vanishes
when $\tau_{\text{sf}} \rightarrow 0$ ($g_{\text{sf}} \rightarrow
\infty $). The spin-accumulation causes a reduction of the
conductance. The resulting conductance (\ref{cond}) can be understood
in terms of the equivalent circuit shown in Fig.\ (\ref{f:equcir}),
which is identical to the high-temperature classical circuit for two
spins,\cite{Brataas99:93} but now generalized to include Coulomb
charging effects by renormalizing the conductnaces so that
$g^l_{\uparrow} \rightarrow g_0 g^l_{\uparrow}$, $g^l_{\downarrow}
\rightarrow g_0 g^l_{\downarrow}$, $g^r_{\uparrow} \rightarrow g_0
g^r_{\uparrow}$ and $g^r_{\downarrow} \rightarrow g_0
g^r_{\downarrow}$. The spin-flip conductance $g_{\text{sf}}$ is not
renormalized, since spin-flip processes do not change the number of
electrons on the island and consequently not the charging energy. We
will now use (\ref{cond}) to find the TMR ratio in F-N-F and F-F-F
SET's.

\section{F-N-F magneto-resistance}
\label{f:FNF}

We first focus on F-N-F systems. The quantity of interest is the
TMR $(g_{\text{P}}-g_{\text{AP}})/g_{\text{P}}$.  When
the magnetizations in the left and right leads are parallel, the
tunnel conductances can be characterized as $g^l_s=g^l(1+sP_l)/2$ and
$g^r_s=g^r(1+sP_r)/2$, where $s=+1$ ($s=-1$) denotes spin-up
(spin-down) and
$P_l=(g^l_{\uparrow}-g^l_{\downarrow})/(g^l_{\uparrow}+g^l_{\downarrow})$
and
$P_r=(g^r_{\uparrow}-g^r_{\downarrow})/(g^r_{\uparrow}+g^r_{\downarrow})$
denote the polarizations of the left and right tunnel junctions,
respectively. In the anti-parallel configuration the tunnel
conductances are $g^l_s=g^l(1+sP_l)/2$ and $g^r_s=g^r(1-sP_r)/2$. The
total conductances of each junction,
$g^l=g^l_{\uparrow}+g^l_{\downarrow}$ and
$g^r=g^r_{\uparrow}+g^r_{\downarrow}$, are thus independent of the
magnetization configuration. Hence, the correlation function $g_0$ and
consequently also the prefactor in (\ref{cond}) remain invariant on
going from the parallel to the anti-parallel configuration. The
magneto-resistive effect is thus solely due to the spin-accumulation
which is contained in the second term of Eq.\ (\ref{cond}).  The
TMR of F-N-F systems is
\begin{equation}
\frac{g_{\text{P}}-g_{\text{AP}}}{g_{\text{P}}} = \frac{P_l
P_r}{1+\alpha^2_g -\beta^2_g} (1-\gamma^2_g) \, ,
\label{MR}
\end{equation}
where the parameter $\alpha^2_g=4g_{\text{sf}} /[g_0 (g^l+g^r)]$
($\alpha^2_g \ge 0$) reduces the spin-accumulation and hence the TMR
due to spin-flip relaxation, $\gamma_g=(g^l-g^r)/(g^l+g^r)$ ($0\le
\gamma^2_g \le 1$) reduces the TMR due to the asymmetries in the
conductances of the left and right junctions, and $\beta^2_g=(P_l^2
(1+\gamma_g)^2 + P_r^2 (1-\gamma_g)^2 - 2P_l P_r (1-\gamma^2_g))/4$
($0 \le \beta^2_g \le 1$). The TMR is proportional to the polarization
of the left and the right tunnel conductance. This result (\ref{MR})
reduced to the high temperature or large bias result in Ref.\
\onlinecite{Brataas99:93} when $g_0 \rightarrow 1$.
 
The spin-flip relaxation time is governed by spin-orbit coupling and
magnetic impurities at low temperatures. We assume that
$\tau_{\text{sf}}$ is temperature-independent below the Coulomb
charging energy (at very low temperatures this assumption can become
invalid, but that is a regime beyond the investigation in this
article).  Firstly, since the correlation function $g_0$ can only
decrease due to Coulomb charging effects the factor $\alpha_g$
appearing in (\ref{MR}) can only increase at low temperatures and {\em
the TMR in the Coulomb blockade regime is equal to or smaller than the
classical high-temperature TMR}. Secondly, when the spin-flip
relaxation time is much larger than the dwell time, the TMR is
independent of the charging energy, temperature and tunnel resistances
for practically all temperatures since $\alpha_g \rightarrow 0$ except
at extremely low temperatures. Thirdly, in this regime we expect the
nonlinear response TMR to have a re-entrant behavior: The linear
response TMR is identical to the TMR when the bias is much larger than
the Coulomb charging energy which equals the classical
high-temperature TMR. In the intermediate bias regime effects of the
Coulomb charging energy
appear\cite{Barnas98:1058,Barnas98:85,Brataas99:93} in the sequential
tunneling limit and an enhancement of the TMR has been demonstrated in
Ref.\ \onlinecite{Imamura99:6017} using the co-tunneling formalism. Finally
(\ref{cond}) and (\ref{MR}) can be generalized to finite frequencies
by letting $g_0 \rightarrow g_0(\omega)$ and $1/\tau_{\text{sf}}
\rightarrow 1/\tau_{\text{sf}} + i\omega$ so that the AC response can
be used to detect the spin-accumulation.\cite{Brataas99:93} Our
general result (\ref{MR}) is consistent with the calculations
performed in the sequential tunneling regime.
\cite{Barnas98:85,Brataas99:93,Korotkov99:89}

\section{F-F-F magneto-resistance}
\label{s:FFF}

The spin-flip relaxation time in a ferromagnet is typically much shorter
than the transport dwell time.\cite{Brataas99:93} In the limit
$\tau_{\text{sf}} \rightarrow 0$ the conductance of F-F-F systems
reduces to\cite{Wang99:5138}
\begin{equation}
g=\frac{g^lg^r}{g^l+g^r} g_0(\beta E_c,g^l+g^r,n_{\text{ex}}) \, .
\end{equation}
The TMR of F-F-F systems is defined as  
\begin{equation}
\gamma
=\frac{R_{\text{AP}}-R_{\text{P}}}{R_{\text{P}}}=\frac{g_{\text{P}}
-g_{\text{AP}}}{g_{\text{AP}}} \,,
\label{tmr}
\end{equation}
where $R_{\text{P}}=R_K/g_{\text{P}}$
($R_{\text{AP}}=R_K/g_{\text{AP}}$) is the resistance of the FSET and
$g_{\text{P}}$ ($g_{\text{AP}}$) is the dimensionless conductance of
the FSET when the magnetizations in the leads and the central island
are parallel (anti-parallel). (Note that we use a slightly different
definition for the relative tunneling magnetoresistance in F-N-F
systems than in F-F-F systems in order to make our discussions
coherent with previous work). For F-F-F systems it is assumed that
the magnetizations in the leads remain parallel, and that only the
magnetization direction of the central island changes by applying an
external magnetic field.

The TMR is caused by the magnetization configuration
dependence of the total junction conductances
$g^l=g^l_{\uparrow}+g^l_{\downarrow}$ and
$g^r=g^r_{\uparrow}+g^r_{\downarrow}$ and therefore also by the
magnetization configuration dependence of the renormalization factor
$g_0$. The latter dependence gives rise to the enhancement of the
TMR in the Coulomb blockade regime. The first term of
the effective action (\ref{wirk}) governing the renormalization of the
conductance $g_0$ (\ref{g0}) only depends on the Coulomb charging
energy of the device, and thus in the weak tunneling limit
$g_0\rightarrow 1$ and the dimensionless conductances and the TMR of
the FSET reduce to the classical values.  The second term in the
action (\ref{wirk}) is proportional to the sum of the dimensionless
conductances of the tunnel junctions and consequently is magnetization
configuration dependent.

In order to evaluate the TMR the conductances of the FSET in different
alignments must be calculated.  We characterize the polarizations of
the junctions by the relative polarizations $P$ and $P'$ associated
with the reservoirs and the island, respectively. In a simple two-band
model with spin-independent tunneling probabilities the polarizations
$P$ and $P'$ are directly related to the density of states in the
reservoirs and the island. In general there is not such a one-to-one
correspondence between the polarizations of the tunnel barriers and
the polarizations of the density of states, but we can still
characterize the spin-dependent asymmetries in the tunnel conductances
by the parameters $P$ and $P'$.  We denote the dimensionless
conductance of the left (right) junction averaged over the parallel
and anti-parallel alignments by $\bar{g}^{l/r}$. In the parallel
configuration the spin-dependent conductances are
\begin{equation}
g^{l/r}_{\text{P},s} = \frac{1}{2} \bar{g}^{l/r} (1+sP)(1+sP') \, ,
\end{equation}
where $s=1$ ($s=-1$) for spin-up (spin-down) electrons. Hence, the
total classical dimensionless junction conductance is
\begin{equation}
g_{\text{P}}^{\text{cl}} =g^l_{\text{P}}
+g^r_{\text{P}} =\bar{g} (1+P P^{\prime }) \, ,
\end{equation}
where $\bar{g}=\bar{g}^{l}+\bar{g}^{r}$.  Similarly, in the
anti-parallel configuration the conductance is
\begin{equation}
g^{l/r}_{\text{AP},s} = \frac{1}{2} \bar{g}^{l/r} (1+sP)(1-sP') \, .
\end{equation}
The total classical dimensionless junction conductance is thus
\begin{equation}
g_{\text{AP}}^{\text{cl}}= g^l_{\text{AP}}+g^r_{\text{AP}} =\bar{g}
(1-P P^{\prime }) \,.
\end{equation}
In the high temperature limit, one can substitute the forms of
$g_{\text{P}}^{\text{cl}}$ and $g_{\text{AP}}^{\text{cl}}$ described
above into the definition of the TMR (\ref{tmr}), and obtain a simple
expression $\gamma^{\text{cl}}=2PP'/(1-PP')$ for the classical TMR.

If the thermal energy is larger than the Coulomb charging energy, the
path integrals can be evaluated semiclassically, and the lowest order
quantum corrections to the TMR can be obtained. In this regime the
conductance of the FSET is \cite{Wang99:5138,Wang97:12868}
\begin{eqnarray}
g_0 (\omega = 0) \simeq & & 1-\frac{\beta E_c}{3}+(0.0667
+0.0185g_{\eta}^{\text{cl}}) (\beta E_c)^2 \nonumber \\ & &
-\frac{8\pi^2}{\beta E_c} e^{-\pi^2/\beta E_c-g_{\eta}^{\text{cl}}/2}
\cos (2\pi n_{{\text{ex}}}) \,
\end{eqnarray}
with $\eta=P (AP)$ for the parallel (anti-parallel) alignment. 
Substituting this expression into the definition of the TMR (\ref{tmr}), 
we obtain 
\begin{eqnarray}
\gamma^{\text{se}} &=& \gamma ^{\text{cl}}+\mu ^{\text{se}}\left(
1+\gamma ^{\text{cl}}\right) \left( \beta E_{c} \right) ^{2} + \delta
\gamma (n_{{\text{ex}}}) \, ,
\label{trse}
\end{eqnarray}
where $\mu^{\text{se}}=0.02 g_{\text{P}}^{\text{cl}}$, and the
leading term of the gate voltage dependent part of the semiclassical
TMR is
\begin{eqnarray}
\delta \gamma (n_{{\text{ex}}}) &=& \frac{8\pi^2}{\beta E_c} (1+\gamma
^{\text{cl}}) e^{-\pi^2/\beta E_c} \times \nonumber \\
& & [e^{-g_{\text{AP}}^{\text{cl}}/2}
-e^{-g_{\text{P}}^{\text{cl}}/2}] \cos (2\pi n_{{\text{ex}}})\, .
\label{trse_delta}
\end{eqnarray}
In Ref.\ \onlinecite{Wang99:5138}, $n_{\text{ex}}=0$ was considered and
consequently the gate voltage dependent contribution
(\ref{trse_delta}) was disregarded.  For large junction conductances
or high temperatures, this term is exponentially suppressed and the
influence of the gate voltage on the TMR is negligible.  The gate
voltage dependence part of the TMR is important at low
temperatures. Although Eq.\ (\ref{trse_delta}) is not accurate at very
low temperatures, it can be used to estimate when $\delta \gamma
(n_{{\text{ex}}})$ becomes relevant.  For FSET's with positive
polarizations $PP'>0$ the conductance in the anti-parallel
configuration is smaller than the conductance in the parallel
configuration $g_{\text{AP}}^{\text{cl}} < g_{\text{P}}^{\text{cl}}$,
and the maximum TMR occurs when $n_{{\text{ex}}}=0$. The TMR attains
its minimum at $n_{{\text{ex}}}=1/2$, and returns to the maximum at
$n_{{\text{ex}}}=1$ since the TMR is a periodic function of
$n_{\text{ex}}$ with period 1.

At low temperatures, an analytical expression of the TMR is not
available. We will study the maximum and minimum value of the TMR at
low temperatures via Monte-Carlo simulations. We use Ni leads with
$P=0.23$, and Co island with $P'=0.35$ in all numerical calculations
below.

\subsection{TMR in the 'off'-state}

In the absence of induced charges, the geometric phase factor $2\pi
ikn_{{\text{ex}}}$ in the phase-phase correlation function (\ref{ppc})
vanishes.  The density matrix for each winding number $k$ is then
positive and the phase-phase correlation function can be computed
directly using standard Monte Carlo simulation techniques.  In our
earlier work\cite{Wang99:5138} a large enhancement of the TMR, roughly
4 times larger than the classical TMR for $\bar{g}=12$ and $E_c/k_B
T=40$, was found.  In this section we will demonstrate that the TMR is
further enhanced at even lower temperatures. We will also address
whether even larger junction conductances increase or decrease the TMR
ratio.

The enhancement of the TMR ratio can be clearly demonstrated in the
co-tunneling regime, where the conductance can be calculated
analytically.\cite{Averin90:2446,Averin91} Since co-tunneling is a
second order tunneling process and the first order sequential
tunneling vanishes in the Coulomb blockade regime, the conductance of
the FSET is proportional to the square of the classical conductance
with a temperature dependent prefactor.  Consequently the TMR
(\ref{tmr}) is enhanced to a value slightly larger than twice the
classical TMR.\cite{Takahashi98:1758} A naive extension of this result
is that beyond the sequential and co-tunneling regimes, the TMR is
enhanced by a factor of $n$ when the $n$-th order tunneling process
dominates.  However this picture is too simplified since contributions
from different orders of tunneling have different
temperature-dependent prefactors and one cannot in general specify the
tunneling process that dominates for given junction resistances and
temperature.  In the strong tunneling regime perturbation theory
breaks down and a number of higher order tunneling processes have to be
taken into account simultaneously.

When the thermal energy is not much smaller than the charging energy,
the semiclassical formula (\ref{trse}) can be modified to fit the TMR
values by replacing the Coulomb charging energy by a renormalized
charging energy.  This substitution agrees well with the results of
earlier investigations on the single-electron
box.\cite{Buttiker87:3548,Wang97:545} The renormalization scheme also
predicts that the effective charging energy at sufficiently low
temperature decreases for larger junction conductances due to the
coupling to the leads.  This means that for a given temperature, the
TMR in the strong tunneling regime is smaller for larger junction
conductances.  The role of strong tunneling is therefore {\em
dual}.\cite{na} The constructive role is to increase the TMR by the
inclusion of higher order tunneling processes at low temperature. The
destructive role is caused by the renormalized charging energy to a
smaller value $E_c^* \ll E_c$ so that the TMR is closer to the
classical value at a given temperature.

When the thermal energy is much smaller than the Coulomb charging
energy strong tunneling is manifested as an enhancement of the TMR by
lowering the temperature. We compute the TMR values for two values of
the classical total dimensionless conductance $\bar{g}=12$ and
$\bar{g}=24$.  For a given conductance $\bar{g}$ and temperature the
phase-phase correlation function (\ref{phasephase}) is computed via
Monte-Carlo simulations for a sufficiently large number of discrete
points in the imaginary time interval $[0, \beta]$.  The phase-phase
correlation functions after two succeeding $N$ samplings are compared.
If the curves are smooth and sufficiently close to each other, they
predict roughly the same conductance after the analytical continuation
of the Matsubara-Fourier components.  The small difference determines
the uncertainty of the final result, which is shown as the size of the
data point in the figures below.  When the deviation between the two
phase-phase correlation functions is larger than the requested
accuracy they are averaged to give the phase-phase correlation
function for the case of $2N$ samplings, and a further simulation is
performed for the next case of $2N$ samplings.  This procedure is
followed until the final result converges.  Using the computed
phase-phase correlation functions, the conductances of the TMR can be
calculated via the Kubo's formula Eq.~(\ref{g0}).  The analytical
continuation from the Matsubara frequency to the real
frequency\cite{Mahan90} is performed via the P\'ade
approximate.\cite{Chakravarty95:501}

The resulting TMR ratio in the absence of the gate voltage is shown in
Fig.\ (\ref{TMRoff}) for $\bar{g}=12$ and 24.  In both cases, the TMR
values at low temperature are larger than the classical values, which
shows that the TMR is enhanced by higher order tunneling processes in
the Coulomb blockade regime.  For $\bar{g}=12$ at $k_B T/E_c=0.01$,
the TMR is enhanced by a factor of 8 compared to the classical value,
which is almost twice as large as the value that we reported for $k_B
T/E_c=0.025$ in our previous work.\cite{Wang99:5138} For a given
temperature, the TMR for $\bar{g}=12$ is larger than the one for
$\bar{g}=24$ in the temperature region shown in Fig.\ (\ref{TMRoff}),
which means that the smearing of the Coulomb charging energy via
higher order tunneling processes is stronger for larger junction
conductances.  Therefore for experimentally easily accessible
temperatures, only moderate enhancement of the TMR for the FSET formed
by tunnel junctions with extremely large conductances can be obtained.
In order to experimentally observe large TMR ratios in SET's tunnel
conductances of the order of 10 times the quantum conductance are
recommended.

\subsection{TMR in the 'on'-state}

In the 'on'-state when $n_{\text{ex}}=C_g V_g/e=1/2$ module $1$, there
is no Coulomb blockade effect even when the thermal energy is much
smaller than the charging energy.  In the sequential tunneling regime,
the conductance of the FSET approaches half of the classical value
when the temperature is lowered.\cite{Joyez97:1349} Beyond the
sequential tunneling regime, the reduction of the conductance of the
FSET by lowering the temperature is slow at relatively high
temperatures, but at sufficiently low temperature the conductance
attains values smaller than the conductance in the sequential
tunneling regime.\cite{Takahashi98:1758,Joyez97:1349} The conductance
in the 'on'-state in the strong tunneling limit and its implication
for the TMR will be investigated in this section.

In order to calculate the 'on'-state conductance in the strong
tunneling regime by Monte-Carlo simulations, we reformulate the
phase-phase correlation function as
\begin{equation}
\Gamma^{{\text{on}}} (\tau )=
\Gamma^{{\text{re}}} (\tau )/Z^{\text{re}} \, ,
\end{equation}
where the function $\Gamma^{\text{re}} (\tau )$ is the mean value of 
the phase factor 
\begin{eqnarray}
\Gamma^{\text{re}} (\tau ) & = & \frac{1}{Z^{0}} \sum_{k=-\infty
}^{\infty } \int_{b_{k}}D\varphi e^{-S[\varphi ]} \cos (\pi k) \times
\nonumber \\
&& \cos [\varphi (\tau )-\varphi (0)]\, ,
\end{eqnarray}
with respect to the partition function $Z^0$ which is a direct
summation of the density matrices for all winding numbers $k$,
$Z^{0}=\sum_{k=-\infty }^{\infty } \int_{b_{k}}D\varphi e^{-S[\varphi
]} $ and the partition function at resonance is renormalized in a
similar way,
\begin{equation}  
Z^{\text{re}}=\frac{1}{Z^{0}} \sum_{k=-\infty }^{\infty } 
\int_{b_{k}}D\varphi e^{-S[\varphi ]} \cos (\pi k) \, . 
\end{equation}
In the above formulas, both the numerator and the denominator of the
phase-phase correlation function $\Gamma^{\text{on}} (\tau )$ can be
calculated with respect to the density matrices of the FSET with a
geometric factor which equals one. The computation can therefore be
performed in the same way as in the case of vanishing gate
voltage. The complication is that we now have to calculate both
$\Gamma^{\text{re}} (\tau )$ and $Z^{\text{re}}$ instead of a single
correlation function.

When the thermal energy is not much smaller than the charging energy,
quantum fluctuations are small and the simulations are rather
fast. Consequently the stable phase-phase correlation functions can be
readily obtained using the approach described above.  A typical result
is shown in Fig.\ (\ref{phasephase2}).  At lower temperatures, the imaginary
time $\beta$ is longer, and more lattice points are needed to
guarantee the convergence of the Trotter product.  Moreover, the
computed correlation functions for the same amount of sampling numbers
turn out to fluctuate more strongly at lower temperatures as shown in
Fig.\ (\ref{phasephase3}).  For $\bar{g}=12$ at $\beta E_c \geq 60$ and
$\bar{g}=24$ at $\beta E_c \geq 100$, the correlation functions
computed via the approach of doubling the sampling number still
fluctuate considerably within a realistic computing time.  Therefore
we do not attempt to perform simulations for temperatures lower than
$\beta E_c=100$.  At the lowest temperature the correlation function
at the points with large fluctuations is replaced by a best fit smooth
function so that the analytical continuations in the Kubo's formula
lead to a converging conductance of the FSET.  The accuracy of this
smoothing approach needs to be determined by long time Monte Carlo
simulations in the future, since we cannot guarantee that the single
data point at the lowest temperature is accurate within its size. 

The TMR ratio as a function of the inverse dimensionless temperature
is shown in Fig.\ (\ref{TMRon}). For $\bar{g}=12$, the TMR first
increases by lowering the temperature.  The values are almost the same
as the case of the vanishing gate voltage.  Around $\beta E_c =10$,
the gate voltage dependent contributions to the TMR becomes important
shifting the TMR value for $C_g V_g=e/2$ farther away from the value
for $C_g V_g=0$ at lower temperatures.  Therefore while the TMR in the
Coulomb blockade regime is enhanced further at lower temperatures, the
TMR at resonance decreases rapidly to the values smaller than the
classical TMR for $\beta E_c \geq 30$.  Negative TMR values could
possibly occur at even lower temperatures, which needs to be
determined quantitatively by further investigations which at present
cannot be performed due to the computing time required.  For
$\bar{g}=24$, in contrast to the case of smaller junction conductances
$\bar{g}=12$, the TMR increases constantly and slowly by lowering the
temperature down to a value of $0.01E_c/k_B$. The curve more or less
follows the tendency of the TMR in the case of vanishing gate voltage.
This is understandable because the gate voltage dependent
contributions to the TMR is proportional to the factor $\exp
(-g_{\eta}^{\text{cl}}/2)$ as expressed in (\ref{trse_delta}), which
is smaller for larger $\bar{g}$, therefore the difference between the
maximal TMR at $C_g V_g=0$ and the minimal TMR at $C_g V_g=e/2$
determined by the gate voltage dependent contribution is also smaller,
and the variation of the minimal TMR will change the direction at
lower temperatures.  In both cases, the difference between the maximal
TMR and the minimal TMR is found to be larger at lower temperatures,
therefore a wider variation range of the TMR can be achieved by tuning
the gate voltage.

\section{Conclusion}
\label{s:conclusion}

We have derived a general formula for the linear conductance of
single-electron transistors containing ferromagnetic elements.  In
F-N-F structures, the TMR is almost independent of the Coulomb
charging energy and is only {\em reduced} at sufficiently low
temperatures when the effective transport dwell time becomes shorter
than the spin-flip relaxation time of the normal-metallic island.  In
F-F-F systems, we have calculated the magneto-resistance as a function
of gate voltages for a wide range of temperatures in the strong
tunneling regime. In the 'off'-state, the magneto-resistance can be
enhanced at low temperatures by higher order tunneling
processes. However, higher order tunneling processes also reduce the
effective charging energy, and slow down the rate of the enhancement
of the magneto-resistance when the junction conductances are much
larger than the quantum conductance.  Consequently the largest
magneto-resistance for a given temperature is obtained when the
junction conductances are larger, but not very much larger than the
quantum conductance.  The magneto-resistance of the device in the 'on'
state has a more complicated temperature dependence.  Due to the
contributions from the induced charges, the magneto-resistance
increases slowly for very large junction conductances. In contrast to
this, the magneto-resistance is reduced well below the classical
magneto-resistance for not very large junction conductances at
sufficiently low temperatures.

We are grateful to G.\ E.\ W.\ Bauer, K.-A.\ Chao, B.\ I.\ Halperin,
J.\ Inoue, Yu.\ V.\ Nazarov, and A.\ A.\ Odintsov for stimulating
discussions. A.\ B.\ would like to thank the Norwegian Research
Council for financial support and X.\ H.\ W.\ acknowledges support
from the Swedish Research Council.

\begin{figure}
\caption{The single-electron transistor. The central island is
connected to the left and the right reservoirs by tunnel
junctions. The electrostatic energy of the island can be tuned by a
gate voltage capacitively coupled to the island.}
\label{f:set}
\end{figure}

\begin{figure}
\caption{The equivalent circuit of the low-bias conductance of the FSET
including the spin accumulation effect of the central island.}
\label{f:equcir}
\end{figure}

\begin{figure}[hbt]
\caption{Tunneling magneto-resistance  for $C_g V_g=0$ as a function of the 
inverse of the dimensionless temperatures in the strong tunneling regime 
for $\protect \bar{g}=12$ (filled dotted), and for $\protect \bar{g}=24$ 
(filled diamonds). The errors are smaller than the sizes of the symbols.
}
\label{TMRoff}
\end{figure}

\begin{figure}[hbt]
\caption{Phase-phase correlation function as a function of the imaginary 
time for $\protect \bar{g}=12$, $\protect \beta E_c=20$ and $C_g V_g=e/2$.  
The dotted lines are the Monte-Carlo data for $10^6$ samplings, each.  The 
dot-dashed line is the average of the dotted lines, the dashed line is the
Monte-Carlo data for next $2 \times 10^6$ samplings, and the solid line 
is the average of the dot-dashed line and the dashed line.
}
\label{phasephase2}
\end{figure}

\begin{figure}[hbt]
\caption{Phase-phase correlation function as a function of the imaginary 
time for $\protect \bar{g}=12$, $\protect \beta E_c=60$ and $C_g V_g=e/2$.  
The dotted lines are the Monte-Carlo data for $10^6$ samplings, each.  The 
dot-dashed line is the average of the dotted lines, the dashed line is the 
Monte-Carlo data for the next $2 \times 10^6$ samplings, and the solid line 
is the 
average of the dot-dashed line and the dashed line.  The smooth long dashed 
line in the figure is used to replace the fluctuating solid line to 
calculate the Matsubara-Fourier components of the correlation functions. 
}
\label{phasephase3}
\end{figure}

\begin{figure}[hbt]
\caption{Tunneling magneto-resistance  for $C_g V_g=e/2$ as a function of 
the inverse of the dimensionless temperatures in the strong tunneling regime 
for $\protect \bar{g}=12$ (filled dotted), and for $\protect \bar{g}=24$ 
(filled diamonds). The errors are smaller than the sizes of the symbols 
except for the point at lowest temperature for each set of parameters, where
the error estimation is not accurate.
}
\label{TMRon}
\end{figure}


\begin{references}

\bibitem{Prinz98:282} G.\ A.\ Prinz, Science {\bf 282}, 1660 (1998);
P.\ M.\ Levy, Sol.\ St.\ Phys.\ {\bf 47}, 367 (1994); M.\ A.\. M.\
Gijs and G.\ E.\ W.\ Bauer, Adv.\ Phys.\ {\bf 46}, 285 (1997); P.\ M.\
Levy and S.\ F.\ Zhang, Curr. Opin. Solid State Mater. Sci. {\bf 4},
223 (1999); A.\ Brataas, Yu.\ V.\ Nazarov and G.\ E.\ W.\ Bauer,
Phys.\ Rev.\ Lett.\ {\bf 84}, 2481 (2000).

\bibitem{Meservey94:173} R.\ Meservey and P.\ M.\ Tedrow, Phys.\ Rep.\
{\bf 238}, 173 (1994).

\bibitem{Fiederling99:787} R.\ Fiederling, M.\ Kelm, G.\ Reuscher, W.\
Ossau, G.\ Schmidt, A.\ Waag and L.\ W.\ Molenkamp, Nature {\bf 402},
787 (1999); Y.\ Ohno, D.\ K.\ Young, B.\ Beschoten, F.\ Matsukura H.\
Ohne and D.\ D.\ Awschalom, Nature {\bf 402}, 790 (1999).

\bibitem{Tsukagoshi99:572} K.\ Tsukagoshi, B.\ W.\ Alphenaar and H.\
Ago, Nature {\bf 401}, 572 (1999).

\bibitem{Moodera95:3273} J.\ S.\ Moodera, L.\ R.\ Kinder, T.\ M.\ Wong
and R.\ Meservey, Phys.\ Rev.\ Lett.\ {\bf 74}, 3273 (1995); J.\ S.\
Moodera and J.\ Nassar and G.\ Mathon, Annu.\ Rev.\ Mater.\ Sci.\ {\bf
29}, 381 (1999).


\bibitem{Ono97:1261} K.\ Ono, H.\ Shimada and Y.\ J.\ Ootuka, J.\
Phys.\ Soc.\ Jpn.\ {\bf 66}, 1261 (1997); L.\ F.\ Schelp, A.\ Fert,
F.\ Fettar, P.\ Holody, S.\ F.\ Lee, J.\ L.\ Maurice, F.\ Petroff and
A.\ Vaures, Phys.\ Rev.\ B {\bf 56}, R5747 (1997); S.\ Sankar, B.\
Dieny and A.\ E.\ Berkowitz, J.\ Appl.\ Phys.\ {\bf 81}, 5512 (1997).

\bibitem{Barnas98:1058} J.\ Barnas and A.\ Fert, Phys.\ Rev.\ Lett.\
{\bf 80}, 1058 (1998).

\bibitem{Takahashi98:1758} S.\ Takahashi and S.\ Maekawa, Phys.\ Rev.\
Lett.\ {\bf 80} , 1758 (1998).

\bibitem{Barnas98:85} J.\ Barnas and A.\ Fert, Europhys.\ Lett.\ {\bf
44}, 85 (1998).

\bibitem{Brataas99:93} A.\ Brataas, Yu.\ V.\ Nazarov, J.\ Inoue and G.\ E.\
W.\ Bauer, Phys.\ Rev.\ B {\bf 59}, 93 (1999); Eur. Phys. J. B {\bf 
9}, 421 (1999).

\bibitem{Korotkov99:89} A.\ N.\ Korotkov and V.\ I.\ Safarov, Phys.\ Rev.\
B {\bf 59}, 89 (1999).

\bibitem{Imamura99:6017} H.\ Imamura, S.\ Takahashi and S.\ Maekawa,
Phys.\ Rev.\ B {\bf 59}, 6017 (1999).

\bibitem{Wang99:5138} X.\ H.\ Wang and A.\ Brataas, Phys.\ Rev.\
Lett.\ {\bf 83}, 5138 (1999).

\bibitem{MacDonald9912391} A.\ H.\ MacDonald, cond-mat/9912392.

\bibitem{Schon90:198} G.\ Sch{\"{o}}n, and A.\ D.\ Zaikin, Phys.\
Rep.\ {\bf 198} , 237 (1990) and references therein.

\bibitem{Wang97:12868} X.\ H.\ Wang, Phys.\ Rev.\ B {\bf 55}, 12 868
(1997).

\bibitem{Averin90:2446} D.\ V.\ Averin and Yu.\ V.\ Nazarov, Phys.\
Rev.\ Lett.\ {\bf 65}, 2446 (1990).

\bibitem{Averin91} D.\ V.\ Averin and K.\ K.\ Likharev in {\it Mesoscopic
Phenomena in Solids}, edited by B.\ L.\ Altshuler, P.\ A.\ Lee, and
R.\ A.\ Webb (Elsevier, Amsterdam, 1991); G.-L.\ Ingold and Yu.\ V.\
Nazarov in {\it Single Charge Tunneling}, Vol.\ 294 of NATO Advanced
Study Institute Series B, edited by H.\ Grabert and M.\ H.\ Devoret
(Plenum, New York, 1992).

\bibitem{Buttiker87:3548} M.\ B{\"{u}}ttiker, Phys.\ Rev.\ B {\bf 36},
3548 (1987).

\bibitem{Wang97:545} X.\ H.\ Wang, R.\ Egger and H.\ Grabert,
Europhys.\ Lett.\ {\bf 38}, 545 (1997); and references therein.

\bibitem{na} We thank Drs.~Korotkov and Nazarov for drawing our attention 
to this point.

\bibitem{Mahan90}  G.\ D.\ Mahan, {\it Many-Particle Physics}, Chapter 3,
(Plenum Press, New York, 1990).

\bibitem{Chakravarty95:501} S.\ Chakravarty and J.\ Rudnick, Phys.\
Rev.\ Lett.\ {\bf 75 }, 501 (1995); and J.\ E.\ Hirsh, in {\it Quantum
Monte Carlo Methods in Equilibrium and Nonequilibrium Systems}, edited
by M.\ Suzuki (Springer, Berlin Heidelberg, 1987).

\bibitem{Joyez97:1349} P.~Joyez, V.~Bouchiat, D.~Esteve, C.~Urbina and
M.~H.~Devoret, Phys.\ Rev.\ Lett.\ {\bf 79}, 1349 (1997).

\end{references}
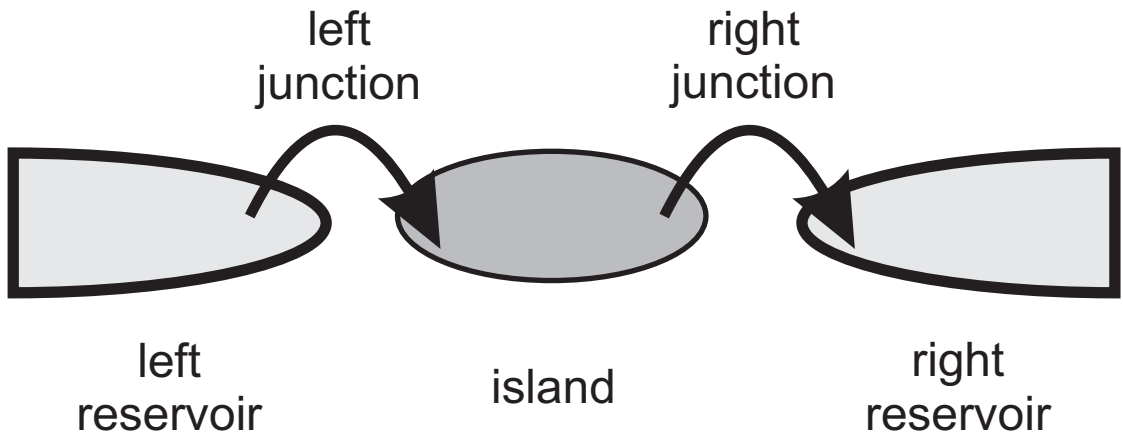
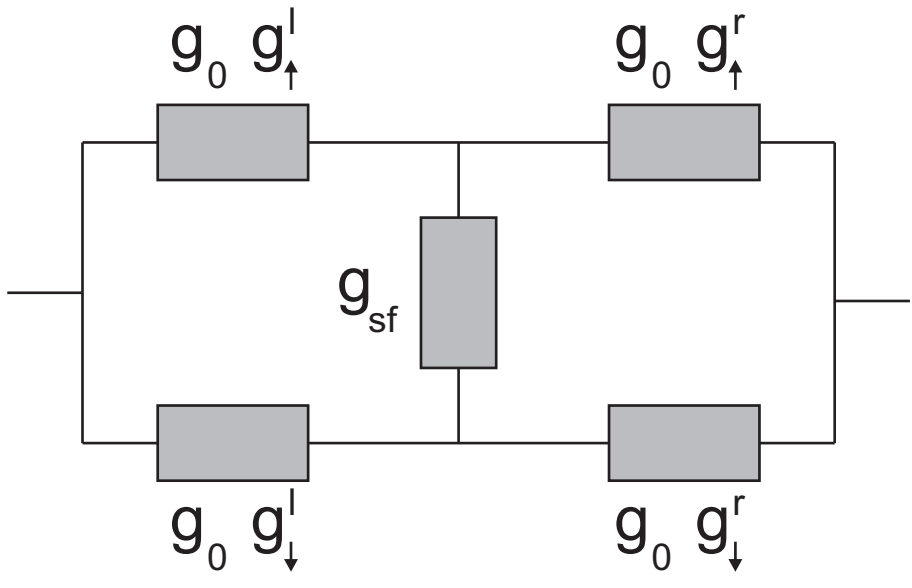
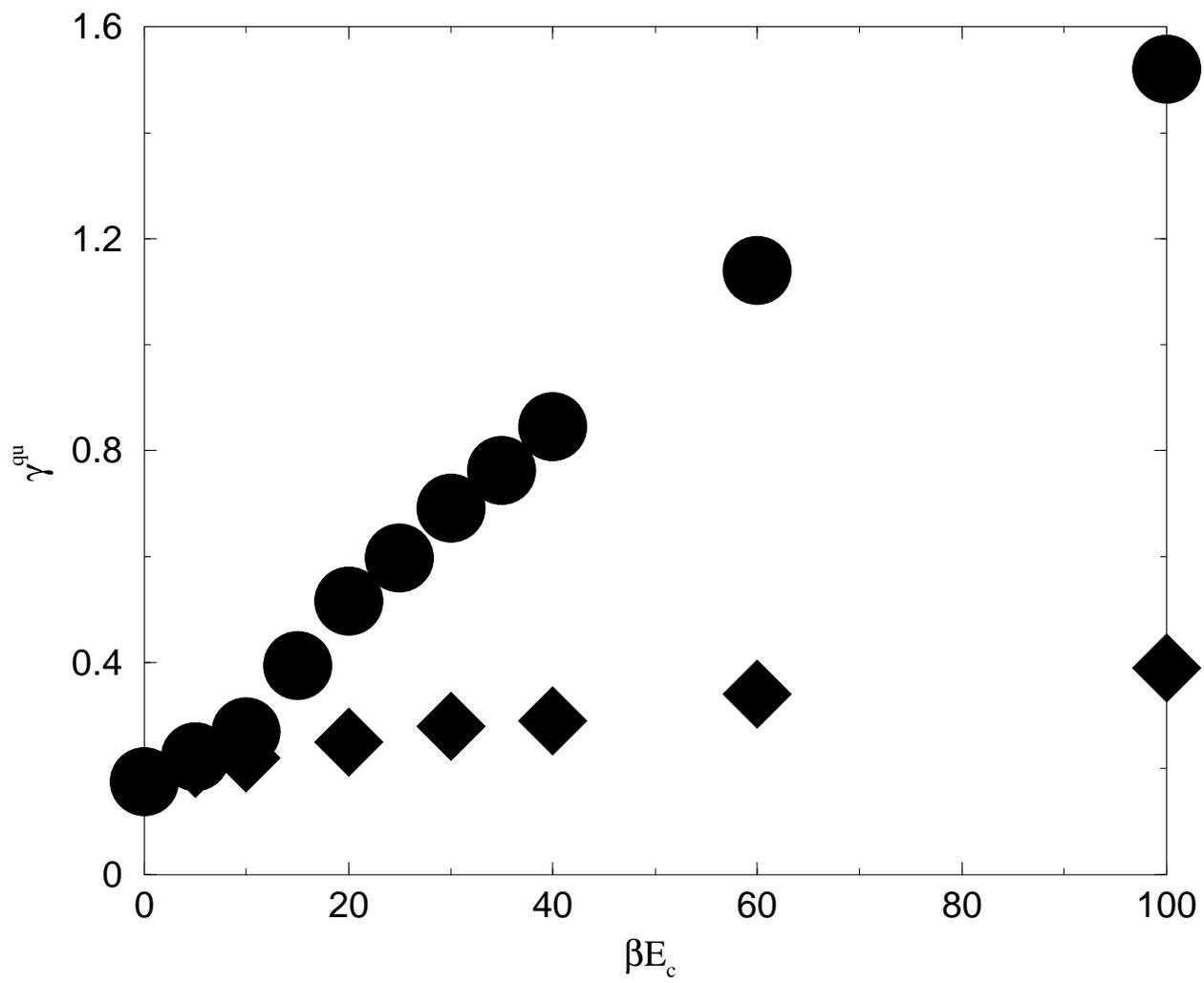
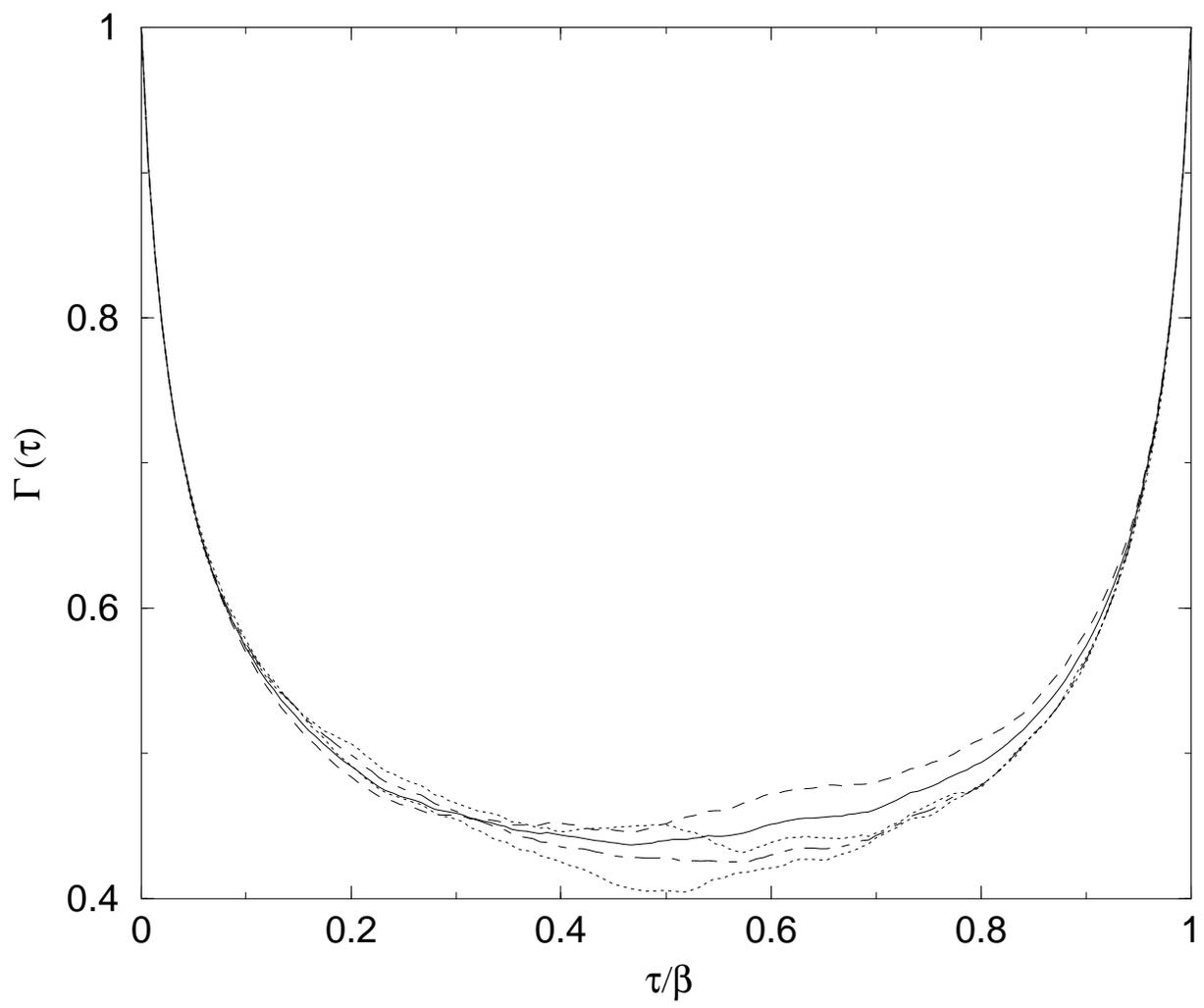
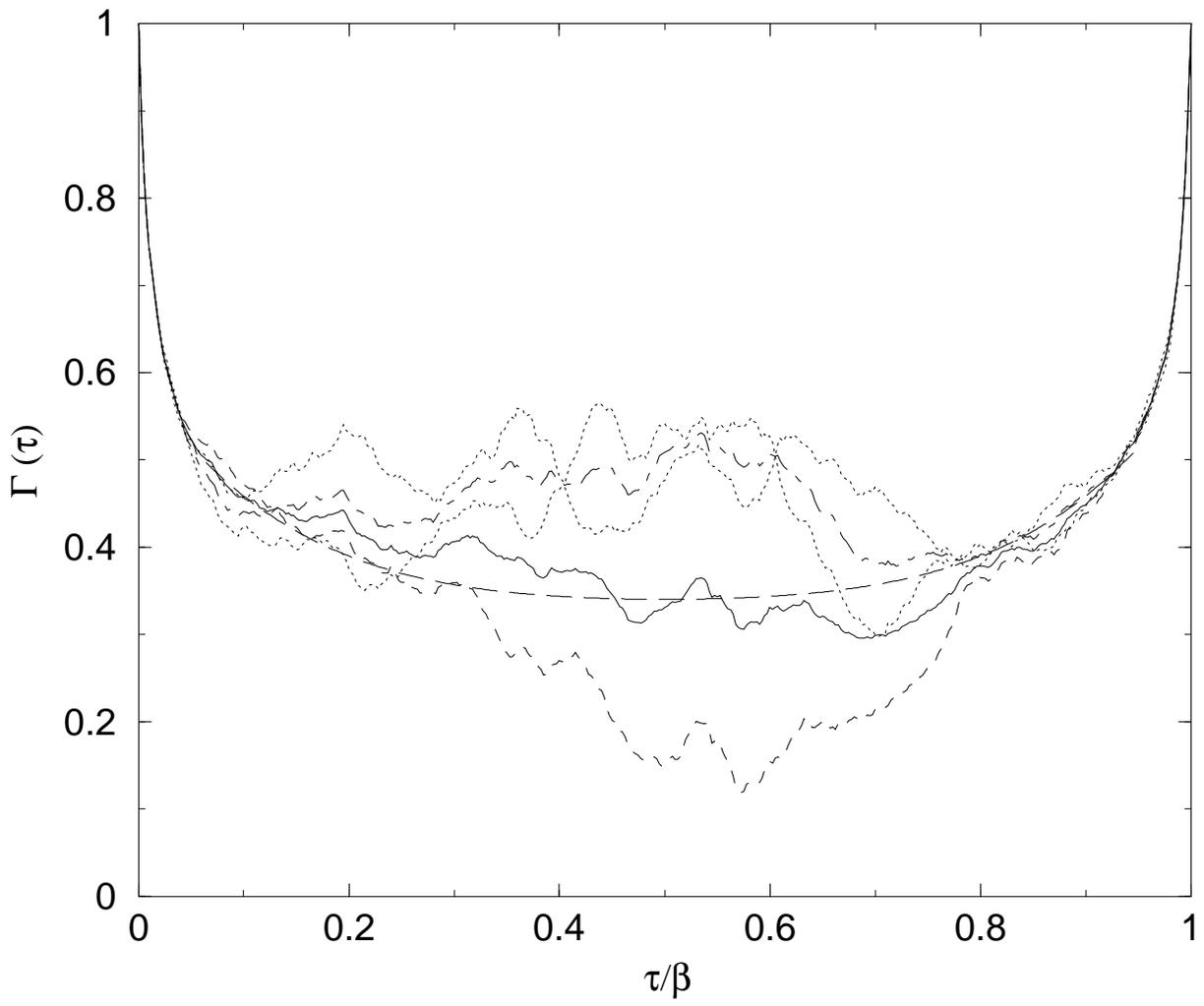
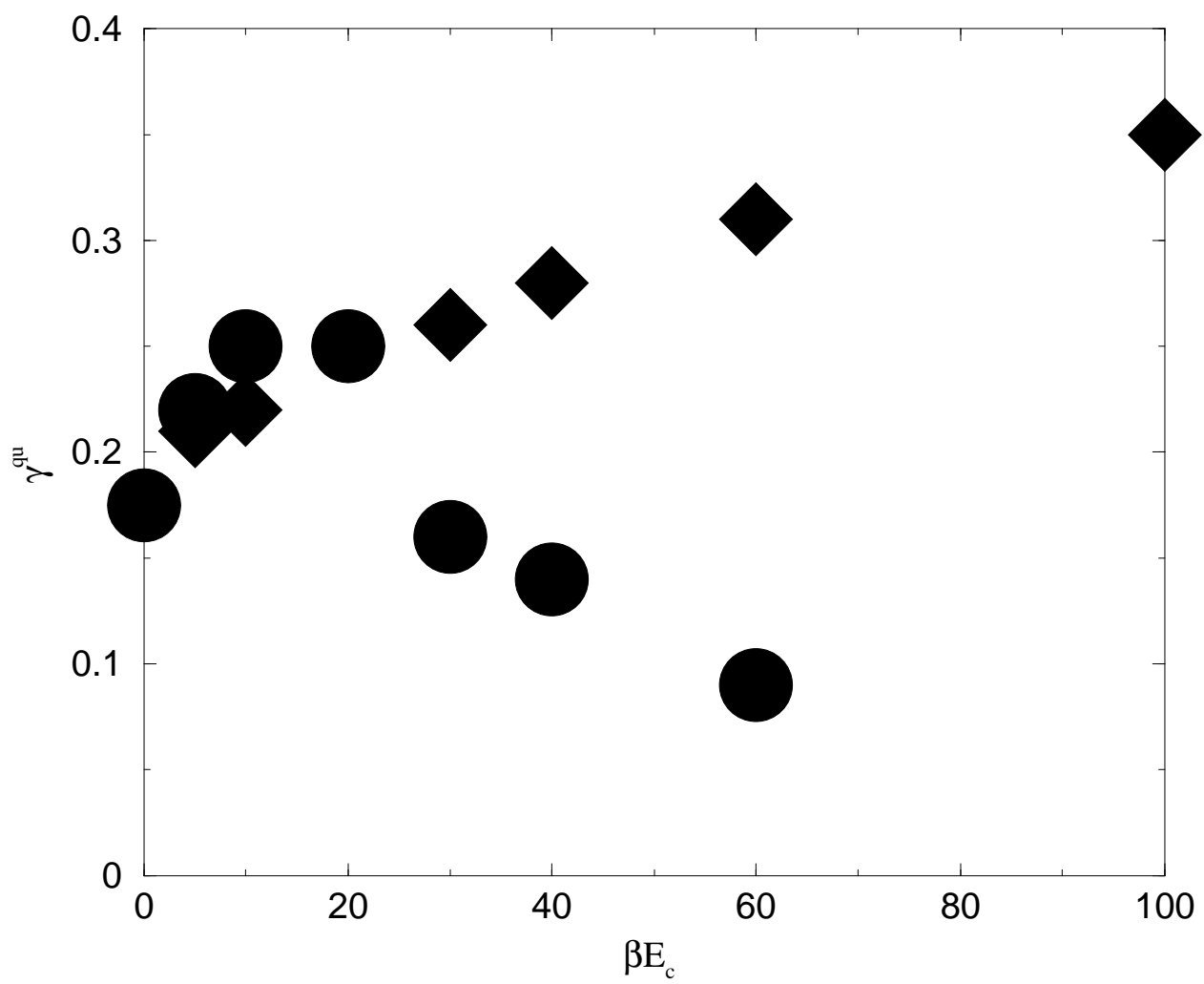
\end{document}